\pdfoutput=1
\documentclass[aps,twoside,twocolumn,pra,superscriptaddress,10pt]{revtex4-2}
\usepackage{epsfig,amsmath,amssymb}
\usepackage{graphicx}
\usepackage{array}
\usepackage{xcolor}
\usepackage[colorlinks=true,linkcolor=blue]{hyperref}

\def\vecb#1{\boldsymbol{#1}}
\def\ket#1{|#1\rangle}
\def\bra#1{\langle#1|}

\def\matr#1#2#3{\langle#1|#2|#3\rangle}
\def\abs#1{\left\lvert#1\right\rvert}

\def\ave#1{\langle#1\rangle}

\def\={\!=\!}
\def\>{\!>\!}
\def\<{\!<\!}
\def\-{\!-\!}
\def\+{\!+\!}

\def\abs#1{\left|#1\right|}

\def\uvo#1{\lq\lq #1\rq\rq}

\def\heff{h_{\mathrm{eff}}}

\begin{document}

\title{Unitary death of Schr{\"o}dinger's cat}

\author{Pavel Str\'ansk\'y}
\email{pavel.stransky@matfyz.cuni.cz}
\affiliation{Institute of Particle and Nuclear Physics, Faculty of Mathematics and Physics, Charles University, V Hole\v{s}ovi\v{c}k\'ach 2, 18000 Prague, Czechia}
\author{Pavel Cejnar}
\email{pavel.cejnar@matfyz.cuni.cz}
\affiliation{Institute of Particle and Nuclear Physics, Faculty of Mathematics and Physics, Charles University, V Hole\v{s}ovi\v{c}k\'ach 2, 18000 Prague, Czechia}
\author{Radim Filip}
\email{filip@optics.upol.cz}
\affiliation{Department of Optics, Faculty of Science, Palack{\'y} University, 17.\,listopadu 12, 77146 Olomouc, Czechia}

\date{\today}

\begin{abstract}
We analyze dynamics of the Rabi model describing interactions of a qubit with a single soft-mode oscillator.
We show that the model with a slightly violated parity generates symmetric Schr{\"o}dinger cat states of the oscillator, which suddenly perish in a spontaneous unitary process similar to the measurement-induced wave-function collapse.
The effect is sufficiently robust to be tested experimentally, e.g., with trapped ions, macroscopic mechanical oscillators or superconducting circuits.
\end{abstract}

\maketitle

\section{Introduction}

One of the most magical achievements of quantum theory is the birth of Schr{\"o}dinger's cat---the prediction of genuinely quantum states that simultaneously exhibit two contradictory classical features \cite{Schr35}.
A cat described by a quantum superposition $\ket{\psi}=(\ket{{\rm dead}}+\ket{{\rm alive}})/\sqrt{2}$ is neither dead, nor alive, but somehow both dead and alive simultaneously, giving rise to a new ontological category of reality.
We know that in case of real macroscopic cats, the ubiquitous decoherence processes would surely prevent formation of true superposition states, but the creation of such states with smaller pieces of matter in the laboratory scale became a common subject of top fundamental and applied research, see, e.g., Refs.\,\cite{Haro03,Wine13,John17,Omra19,Girv19,Hack19,Grav23,Bild23,Gupt24}.

An even more intriguing problem follows from the question how Schr{\"o}dinger's cat dies.
This is a metaphor for the so far not fully understood problem of the quantum-to-classical transition, and particularly for the issue of quantum measurement  \cite{Pere95,WhZu83}.
In sharp contrast to the spontaneous evolution of a quantum system, which is smooth and deterministic, the measurement process is rendered as probabilistic sudden reduction (collapse) of the system's state to the one that actually happened to appear in the measurement.
Many theories and interpretations have been developed to bridge the gap between these contradictory concepts of quantum dynamics.
Some approaches propose to interpret the sudden process of quantum state reduction as just an effective description of the unitary measurement process involving quantum entanglement with the measuring agent or an uncontrollable environment---see, e.g., Refs.\,\cite{Ever57,Grif84,Omne92,Zure03,Alla13,Dros18}.   
Some other approaches consider the collapse to be a spontaneous real physical process resulting from various extensions of the present understanding of quantum dynamics---see, e.g., Refs.\,\cite{Ghir86,Brod06,Bass13,Carl22,Penr96,Penr14}.

This article shows that even a purely quantum system, whose unitary dynamics produces Schr{\"o}dinger-cat states (in fact, similar systems are used to generate these states in practice), may also spontaneously (without interactions with an external probe or environment) cause their sudden collapse.
Almost discontinuous changes that appear in the dynamics of our model governed solely by the Schr{\"o}dinger equation can be interpreted as abrupt transformations of Schr{\"o}dinger's cat from the state with equal probabilities of dead and alive components to a state in which one of the alternatives prevails, in analogy to the true measurement-induced reduction of the cat state.
We do not claim to develop yet another solution to the quantum measurement problem; no such solution is actually delivered!
We just intend to demonstrate that even continuous unitary dynamics of a simple system, which is by no means exotic or pathological, may result in phenomena strongly resembling real measurement processes.

\section{Dynamical parity violation}

\begin{figure}[t]
\includegraphics[width=\linewidth]{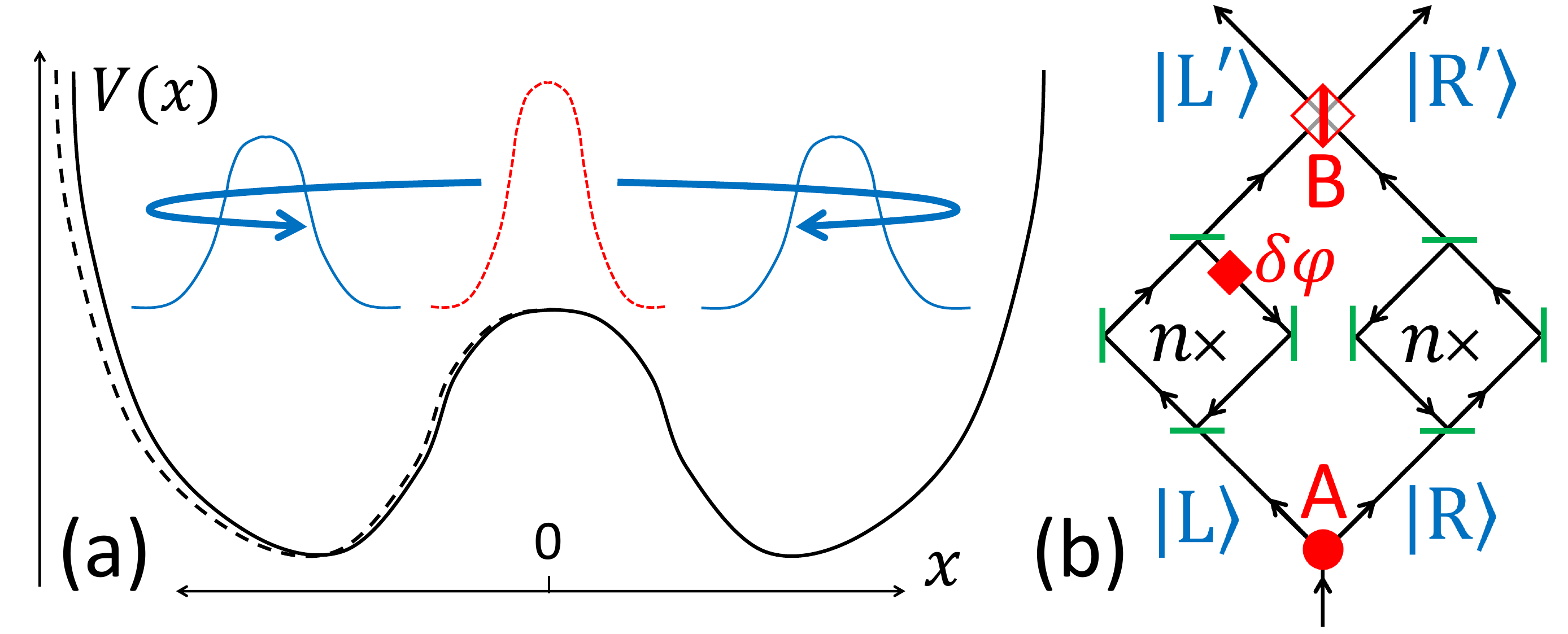}
\caption{A sketch of (a) the parity-violating double well system (see the text) and (b) its optical analog. In panel (b), a single photon passing through the interferometer is in a symmetric cat-like state ${\ket{\psi}=(\ket{{\rm L}}+\ket{{\rm R}})/\sqrt{2}}$ after the beam splitter~A. Each arm of the interferometer contains a loop (with controllable mirrors) in which the photon cycles $n$ times, and in the left arm each cycle induces a small phase shift $\delta\varphi$. After passing the half-reflecting mirror B, the photon is in a generally non-symmetric state ${\ket{\psi'}\propto(\cos\alpha)\ket{{\rm L'}}+(\sin\alpha)\ket{{\rm R'}}}$, where ${\alpha=(n\,\delta\varphi+\frac{\pi}{2})/2}$.} 
\label{schemes}
\end{figure}

The effect we are going to discuss relies on a small violation of parity of the system used to generate the cat state. 
Its essence can be explained on a single particle moving in a slightly asymmetric double-well potential, see Fig.\,\ref{schemes}(a).
The difference between the left and right wells is so marginal that the parity is {\em almost} conserved. 
The particle in a pure state with a fixed parity is put on the top of the central barrier separating both wells.
Any positive- or negative-parity wave function $\psi(x)\={\abs{\psi(x)}e^{i\varphi(x)}}$ yields zero expectation value of momentum, $\matr{\psi}{\hat{p}}{\psi}=\int dx\,\varphi'(x)|\psi(x)|^2$, where $\varphi'(x)$ is the derivative of phase (we set ${\hbar=1}$).
Hence the initial wave packet splits symmetrically into the parts located in both wells.
Each of the split parts propagates to the other side of the respective well and back to the origin.
As the wells are not exactly the same, the parity of the evolving state is being slowly but progressively violated, affecting particularly the coordinate-dependent phase differences between the corresponding parts of the wave function.
However, the probabilities of detecting the particle in the left and right wells remain the same, similarly as probabilities of the dead and alive states of Schr{\"o}dinger's cat.
This lasts until the travel through both wells is completed and all components of the wave function merge together around the origin.
Then the accumulated phases are such that the momentum average is no more zero, which makes the next splitting of the wave function into both wells asymmetric.
The parity is apparently violated, which in some cases leads to a full collapse of the cat state to only one of the alternatives.

The phenomenon pictured above is rooted in the gradual increase of the asymmetry of the cat wave function, which for long remains hidden mostly in phases and only after the merge of the wave-function components generates a large apparent change of probabilities for further evolution.
A similar effect can be achieved in a variety of different ways, e.g., using purely linear optics (with coherent or single-photon states) in the simple device sketched in Fig.\,\ref{schemes}(b).
Here, the photon beam splits symmetrically into components propagating in two arms of an interferometer. 
The phase in one of the arms is shifted $n$~times by a small angle $\delta\varphi$, which leads to apparent asymmetry on the exit.
The asymmetry is maximal for ${n\,\delta\varphi=(2k\!+\!1)\frac{\pi}{2}}$, when all photons appear in the left or right exit for $k$ odd or even integer, respectively.

\section{Qubit-oscillator system}

In the following, we will analyze this effect using an extended Rabi model \cite{Rabi36,Dick54,Jayn63} interpreted as a model describing a single qubit coupled to a single soft-mode scalar field (an oscillator) \cite{Pueb16,Stra21}.
Laboratory realizations of this system with the aid of trapped ions \cite{Lo15,Kien16,Fluh18,Fluh19} or superconducting circuits \cite{Touz19,Camp20,Stas20,Hast21a, Hast21b,Eick22,Siva22,Wang23,Pan23} make it possible to generate Schr{\"o}dinger-cat states and also provide further workable ingredients for quantum technologies \cite{Park17,Park18,Stef19,Chen21}.
If~$\hat{b}^{\dag}$ and~$\hat{b}$ are creation and annihilation operators of the bosonic oscillator quanta and ${\hat{\sigma}_{\pm}=\hat{\sigma}_{x}\pm i\hat{\sigma}_{y}}$, ${\hat{\sigma}_{0}=\hat{\sigma}_{z}}$ are Pauli matrices in the qubit space, a general Rabi Hamiltonian in energy units $\omega$ of the oscillator quantum reads 
\begin{eqnarray}
\hat{{\cal H}}&\equiv&\frac{\hat{H}}{\omega}=\hat{b}^{\dag}\hat{b}+\frac{R}{2}\,\hat{\sigma}_0
+\lambda\sqrt{R}\biggl[\frac{1\!+\!\delta}{2}\left(\hat{b}^{\dag}\hat{\sigma}_{-}\!+\!\hat{b}\hat{\sigma}_{+}\right)
\label{Ham}\\
&&+\frac{1\!-\!\delta}{2}\left(\hat{b}^{\dag}\hat{\sigma}_{+}\!+\!\hat{b}\hat{\sigma}_{-}\right)\biggr]+\mu\sqrt{R}\left(\hat{b}^{\dag}\!+\!\hat{b}\,\right)(\hat{\sigma}_0\!+\!1).
\nonumber
\end{eqnarray}
Here $\lambda$ and $\mu$ are strengths of two independent interaction terms, while ${\delta\in[-1,+1]}$ changes the character of the first interaction between Janes-Cummings (${\delta=+1}$), Dicke (${\delta=0}$) and anti-Janes-Cummings (${\delta=-1}$) regimes \cite{Dick54,Jayn63}.
We assume that self-energies of qubit states $\pm\omega_0/2$ are much larger than $\omega$, so ${R=\omega_0/\omega\gg 1}$.
The dimensionless quantity~$R$, which represents a size parameter of the system \cite{Cejn21}, enters both interaction terms to ensure their proportional scaling with respect to the free Hamiltonian \cite{Stra21}.

We adopt the spin notation for the qubit states, with relations ${\hat{\sigma}_0\ket{\!\uparrow}=+\ket{\!\uparrow}}$, ${\hat{\sigma}_0\ket{\!\downarrow}=-\ket{\!\downarrow}}$, and the occupation number representation of the oscillator states, ${\hat{b}^{\dag}\hat{b}\,\ket{n}=n\ket{n}}$ with ${n=0,1,2,...}$.
The quantum evolution of the whole system is then written in the form
\begin{equation}
\ket{\Psi(t)}\!=\!e^{-i\hat{H}t}\ket{\!\downarrow}\ket{0}=\!\sum_n\bigl[\alpha_{\downarrow n}(t)\ket{\!\downarrow}\!+\!\alpha_{\uparrow n}(t)\ket{\!\uparrow}\bigr]\ket{n},
\label{psit}
\end{equation}
where ${\ket{\!\downarrow}\ket{0}=\ket{\Psi(0)}}$ is the factorized initial state, the ground state of the free Hamiltonian in Eq.\,\eqref{Ham}, and $\alpha$'s are normalized time-dependent amplitudes to be determined in a numerical solution.
State \eqref{psit} for ${t>0}$ is generally entangled in the qubit and oscillator components. 
The oscillator density matrix is obtained by partial tracing over the qubit space, which yields
\begin{equation}
\hat{\rho}_{\rm osc}(t)=\sum_{n,n'}\bigl[\alpha_{\downarrow n}(t)\alpha^*_{\downarrow n'}(t)\!+\!\alpha_{\uparrow n}(t)\alpha^*_{\uparrow n'}(t)\bigr]\ket{n}\bra{n'}.
\label{mad}
\end{equation}

If $R$ is sufficiently large, the average number of bosons in state \eqref{mad} is likely to satisfy ${\ave{\hat{n}}\gg 1}$ and the oscillator can be treated quasi-classically \cite{Stra21}.
Defining effective coordinate and momentum variables of the oscillator,
\begin{equation}
{x=\frac{1}{\sqrt{2R}}\left(\ave{\hat{b}^{\dag}}+\ave{\hat{b}}\right)},\quad 
p={\frac{i}{\sqrt{2R}}\left(\ave{\hat{b}^{\dag}}-\ave{\hat{b}}\right)},
\label{xp}
\end{equation}
where $\ave{\hat{b}^{\dag}}$ and $\ave{\hat{b}}$ stand for expectation values of the oscillator operators, we cast Hamiltonian \eqref{Ham} in the form 
\begin{equation}
\frac{\hat{{\cal H}}(x,p)}{R}=\frac{x^2+p^2}{2}-\vecb{B}(x,p)\cdot\hat{\vecb{\sigma}},
\label{HamB} 
\end{equation}
which acts in the qubit space.
The vector ${\vecb{B}\equiv\vecb{B}(t)}$ can be treated as an external magnetic field which varies as the oscillator state evolves.
If $\ket{\Psi(0)}$ does not strongly overlap with eigenstates of $\hat{H}$ at very high energies (${\sim R\omega}$), the approximate qubit state is $\ket{\!\uparrow_{\vecb{B}(t)}}$, corresponding to spin pointing in the direction of $\vecb{B}(t)$. 
Hence the qubit-oscillator system becomes effectively factorized and Eq.\,\eqref{HamB} yields an effective Hamiltonian
\begin{equation}
\frac{{\cal H}_{\rm eff}(x,p)}{R}=\frac{x^2+p^2}{2}-\abs{\vecb{B}(x,p)},
\label{Hameff}
\end{equation}
which governs the evolution of the oscillator.
We note that the second term in Eq.\,\eqref{Hameff} enriches the simple oscillator dependence in the first term.
On the other hand, if $\ket{\Psi(0)}$ overlaps with high-energy eigenstates, the orthogonal spin component $\ket{\!\downarrow_{\vecb{B}(t)}}$ opposite to $\vecb{B}(t)$ becomes relevant and the factorization assumption fails.

Hamiltonian \eqref{Ham} with ${\mu=0}$ conserves the $Z_2$ symmetry whose parity operator is defined by 
\begin{equation}
\hat{\Pi}=(-1)^{\hat{b}^{\dag}\hat{b}+(\hat{\sigma}_0+1)/2}.
\label{parit}
\end{equation}
The effective Hamiltonian \eqref{Hameff} for ${\mu=0}$ has a single- or double-well form symmetric under the ${(x,p)\leftrightarrow(-x,-p)}$ reflection.
The point ${(x,p)=(0,0)}$, which corresponds to the oscillator initial state~$\ket{0}$, represents the single-well minimum (for ${\lambda\leq \frac{1}{2}}$) or a stationary point (saddle point for ${\frac{1}{2}<\lambda\leq\frac{1}{2|\delta|}}$ or local maximum for ${\lambda>\frac{1}{2|\delta|}}$) separating both ${x<0}$ and ${x>0}$ wells of ${\cal H}_{\rm eff}(x,p)$.
For ${\mu\neq 0}$, the effective Hamiltonian is reflection asymmetric and generally more complicated, but  if $|\mu|<\frac{1}{2}$, the double-well form (for ${\lambda>\frac{1}{2}}$) remains valid.
For more details see Ref.\,\cite{Stra21} and Appendix~\ref{he}.

\section{Schr{\"o}dinger-cat state dynamics}

\begin{figure}[t]
	\includegraphics[width=\linewidth]{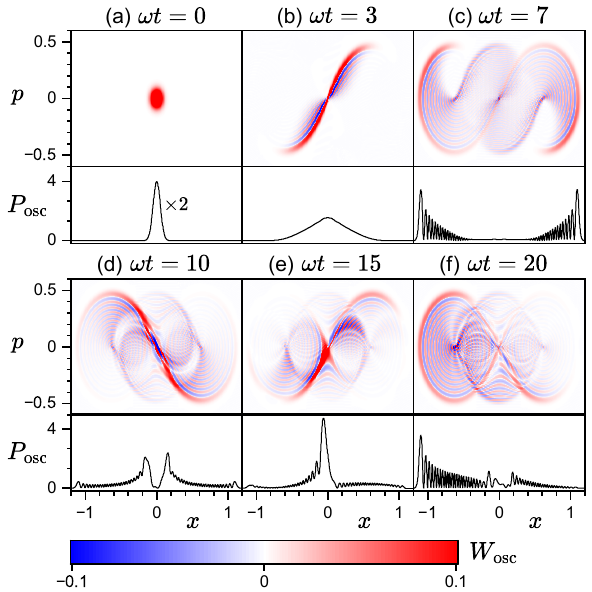}
	\caption{Six snapshots (at times indicated in each panel) of the oscillator Wigner distribution $W_{\rm osc}(x,p,t)$ and the corresponding coordinate distribution $P_{\rm osc}(x,t)$ evolving by the Rabi Hamiltonian \eqref{Ham} with ${R=10^2}$, ${\lambda=0.75}$, ${\delta=0.5}$, and ${\mu=1.3\cdot10^{-3}}$. The initial state $\ket{\!\!\downarrow}\ket{0}$ located around the ${(x,p)=(0,0)}$ saddle point (panel~a) splits symmetrically into two parts traveling in both wells (panels~b--d) until they reunion at the origin. The subsequent new splitting to both wells (panels~e--f) is strongly asymmetric.} 
	\label{wigner}
\end{figure}

Figure \ref{wigner} shows an evolving state of the oscillator in the form of its Wigner quasiprobability distribution in the phase space \cite{Hill84,Polk10}, 
\begin{equation}
W_{\rm osc}(x,p,t)=\frac{1}{\pi}\int_{-\infty}^{+\infty} d\xi\matr{x\!+\!\xi}{\hat{\rho}_{\rm osc}(t)}{x\!-\!\xi}e^{-2ip\xi},
\label{wosc}
\end{equation}
and the corresponding probability distribution of the coordinate, 
\begin{equation}
{P_{\rm osc}(x,t)=\matr{x}{\hat{\rho}_{\rm osc}(t)}{x}=\int  dp\,W_{\rm osc}(x,p,t)}.
\label{posc}
\end{equation}
The calculation follows the exact dynamics of the oscillator state~\eqref{mad} for ${R=10^2}$.
Parameters~$\lambda$ and~$\delta$ are chosen so that the effective Hamiltonian has the double-well form with ${(x,p)=(0,0)}$ being a saddle point. 
Parameter~$\mu$ is set so that the ${x\leftrightarrow-x}$ symmetry is slightly violated.
The initial state from Eq.\,\eqref{psit} has positive parity and the average energy ${\ave{\hat{{\cal H}}}=-\frac{R}{2}}$ equal to the energy of the saddle point.
The figure demonstrates a spontaneous birth of the Schr{\"o}dinger-cat state with the Wigner and probability distribution spit to two equal-weight components traveling in both wells, in analogy to the recent experiment \cite{Bild23}.
The coherence of the superposition is manifested by a pattern of alternating positive and negative values of $W_{\rm osc}(x,p,t)$ and by the oscillatory dependence of $P_{\rm osc}(x,t)$.
When the cat components merge again in the region around the origin, the accumulated parity violations make the next splitting apparently asymmetric, testifying the death of the Schr{\"o}dinger-cat state.
We note that a similar effect is observed for any ${|\delta|<1}$ with~$\lambda$ such that ${\cal H}_{\rm eff}(x,p)$ has the double-well form with a saddle point at the origin.
In contrast, for ${\mu=0}$ the splitting is always symmetric \cite{Stra21}.

\begin{figure}[t]
\includegraphics[width=\linewidth]{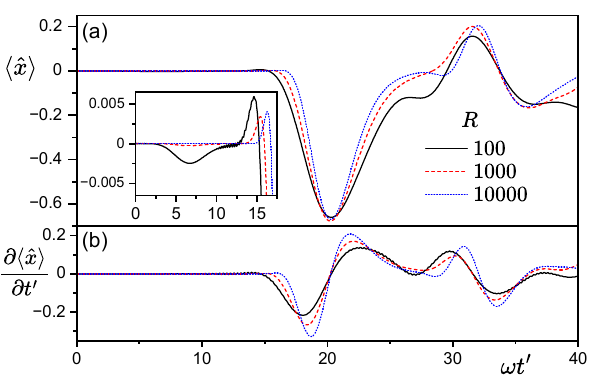}
\caption{Average oscillator coordinate $\ave{\hat{x}(t')}$ (panel a) and its time derivative (panel b) as a function of a rescaled time~$t'$ for evolution~\eqref{psit} with several values of $R$ (individual curves). 
The main plots depict the evolution over about 3 cycles of the split wave function in both wells; the inset expands the first cycle. 
Scaling ${t'=t/s}$ (with ${s=1}$, 1.23, 1.46 for $R={10^2,10^3,10^4}$, respectively) is set so that the main minima of $\ave{\hat{x}(t')}$ coincide. Parameters: ${\lambda=0.75}$, ${\delta=0.5}$, ${\mu=0.13/R}$. } 
\label{avex}
\end{figure}

The suddenness of the death event is studied in Fig.\,\ref{avex}.
Here we compare evolutions of the average coordinate 
\begin{equation}
\ave{\hat{x}(t)}=\frac{{\rm Tr}\bigl[(\hat{b}^{\dag}\!+\!\hat{b})\,\hat{\rho}_{\rm osc}(t)\bigr]}{\sqrt{2R}}
=\iint dx\,dp\,x\,W_{\rm osc}(x,p,t)
\label{xave}
\end{equation}
from the value ${\ave{\hat{x}(0)}=0}$ corresponding to the initial state in Eq.\,\eqref{psit} for different values of parameter $R$.
Since the variation of $R$ modifies the time scale of evolution and the size of parity violation, the time~$t'$ and parameter~$\mu$ were scaled with $R$ to ensure a maximal overlap of individual curves.
In all cases, we see a relatively flat dependence ${\ave{\hat{x}(t')}\approx 0}$ during the first cycle of the split wave packet in both wells.
The little lowering of $\ave{\hat{x}(t')}$ observed in this time domain for smaller values of $R$ is due to the asymmetry of wells.
In  Fig.\,\ref{wigner}, the merge of both wave packet components happens between ${\omega t=10}$ and 15, when the overlap of the evolving and initial states reaches the first maximum (this recurrence time is connected with properties of the local density of states, i.e., distribution of the initial state in the Hamiltonian eigenbasis; see, e.g., Ref.\,\cite{Kloc18}).
In the corresponding narrow time interval we observe a small temporary increase of $\ave{\hat{x}(t')}$ to positive values followed by a sharp drop to deep negative values.
Let us point out that a very similar parity-violating effect, just with the opposite sign, is observed in the qubit expectation value $\ave{\hat{\sigma}_x(t)}$.

The rescaling of time ${t'=t/s}$  in Fig.\,\ref{avex} reflects the effect of localization of the initial state near the stationary point ${(x,p)=(0,0)}$.
The localization increases with $R$, which prolongs the time needed for splitting the wave function into two parts and their reunion after the first cycle.
The stability analysis of dynamics near the stationary point outlined in Appendix~\ref{sc} results in the expression $s=1+A\ln\frac{R'}{R}$, where  ${A}$ is a positive constant and values $R'$ and $R$ correspond to time scales $t'$ and $t$, respectively. 
If watched in the rescaled time, the parity-breaking effect becomes sharper with increasing $R$.
This is verified by studying the rescaled-time derivatives of $\ave{\hat{x}(t')}$ in Fig.\,\ref{avex}(b).

\begin{figure}[t]
\includegraphics[width=\linewidth]{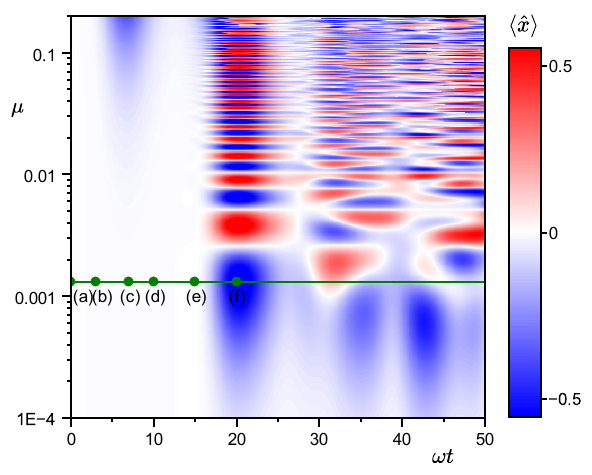}
\caption{Full image of the average oscillator coordinate $\ave{\hat{x}(t)}$ as a function of $t$ and $\mu$ for ${R=10^2}$, ${\lambda=0.75}$, ${\delta=0.5}$.
The red and blue areas mark domains in which $P_{\rm osc}(x,t)$ is located mostly in the right and left well, respectively.
The horizontal line with dots correspond to snapshots (a)--(f) from Fig.\,\ref{wigner}.} 
\label{map}
\end{figure}

Also worth mentioning is the observation of tiny oscillations of some of the curves with smaller values of~$R$ in Fig.\,\ref{avex}. 
These are due to the presence of a low-amplitude component containing the opposite spin orientation $\ket{\!\downarrow_{\vecb{B}(t)}}$ in the full state $\ket{\Psi(t)}$.
The oscillator state associated with this spin orientation is evolved by a different (single-well) effective Hamiltonian ${\cal H}'_{\rm eff}(x,p)$~\cite{Stra21} and yields faster dynamics (speeding up with increasing~$R$) than the dominant component with $\ket{\!\uparrow_{\vecb{B}(t)}}$.
The small size of these oscillations confirms the approximate validity of the above factorization assumption.
We see that the oscillations fade away for very large~$R$.  

Figure~\ref{map} maps the evolving expectation value $\ave{\hat{x}(t)}$ within the time interval of approximately 3.7 full cycles of the split wave function for the parity-breaking parameter taking values $\mu\in[10^{-4},0.2]$.
Parameters $R$, $\lambda$ and $\delta$ are the same as in Fig.\,\ref{wigner}.
The picture for ${\mu<0}$ would be obtained by performing simultaneous inversions ${\mu\to-\mu}$ and ${\ave{\hat{x}(t)}\to-\ave{\hat{x}(t)}}$.
For $\mu\lesssim 10^{-4}$, no abrupt effect after the merge of both Schr{\"o}dinger-cat components occurs, while for $\mu\gtrsim 10^{-1}$ the parity violation becomes obvious already before the merge.
The studied sudden parity-violation effect is observed only between these limits, where it shows very sensitive dependence on $\mu$.
Even a tiny change of the value or sign of $\mu$ may cause a flip of the merged state between the left and right exit channels, i.e., symbolically, alter the collapsed Schr{\"o}dinger's cat between states \uvo{dead} and \uvo{alive}.
This fulfills another essential requirement (besides the suddenness) set upon the process of wave-function collapse---the randomness.
Here, of course, the randomness is not fundamental (in fact, the actual exit channel is deterministic and shows periodic alternation with increasing~$\mu$) but only effective, resulting from imperfect control of the small parameter~$\mu$ in a realistic experimental setup.

\section{Robustness against dissipation}

Aperiodic oscillations of $\ave{\hat{x}(t)}$ seen in Fig.\,\ref{map} continue to infinite times.
Indeed, because of unitary dynamics involving a discrete spectrum of frequencies, the system permanently exhibits approximate recurrences of all states already passed, including the symmetric Schr{\"o}dinger cat state of the first cycle.
However, this will not be so when the system becomes dissipative.
Here we probe the robustness of the sudden parity breaking effect under perturbations induced by interactions of the qubit-oscillator system with surrounding environment.

Because of the essential role of coherence, the effect is supposed to gradually fade away with increasing coupling to the environment.
To provide at least a basic quantification of this prediction, we apply the Lindblad formalism \cite{Brau02} with the master equation containing just a single dissipative term expressed through the jump operator ${\hat{L}=\hat{b}\sqrt{\gamma/R}}$, where the constant~$\gamma/R$ represents the environment-induced damping rate (properly scaled with $R$).
This dissipator captures the dominant damping effect on the oscillator \cite{Haro03,Zure03}.

\begin{figure}[t]
\includegraphics[width=\linewidth]{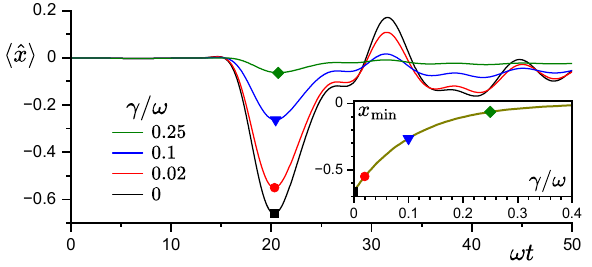}
\caption{Evolution of the average oscillator coordinate in presence of dissipation. 
The main plot depicts $\ave{\hat{x}(t)}$ obtained from the Lindblad equation with the jump operator ${\propto\hat{b}}$ for various damping rates $\gamma/R$, the inset shows the depth of the main minimum as a function of~$\gamma$.
 Parameters as in Fig.\,\ref{avex}.} 
\label{dis}
\end{figure}

The resulting evolution of the average coordinate is shown in Fig.\,\ref{dis}.
In agreement with the above anticipation, the depth of the main minimum of $\ave{\hat{x}(t)}$ decreases with increasing $\gamma$.
Nevertheless, the sudden parity breaking effect remains well visible for $\gamma\lesssim\omega/10$, which provides room for its experimental verification.

\section{Conclusion}

We have revealed an effect of strong dynamical parity breaking in the unitary evolution of the Schr{\"o}dinger cat state governed by a slightly parity-asymmetric Rabi Hamiltonian.
The parity-breaking event is sudden, although not really sharp, and its result is random, although not really indeterministic.
We have emphasized the parallels with the measurement-induced collapse of wave function.
The effect has sufficient resistance against environmental perturbations, which indicates its experimental testability by means of trapped ions \cite{Kien16}, macroscopic mechanical oscillators \cite{Bild23} or superconducting circuits \cite{Pan23}.

More generally, the discussed method of qubit-oscillator coupling with a large value of the energy ratio $R$ represents a promising tool for unitary control of the infinite-dimensional quantum oscillator by a single ancillary qubit.
This tool can find a plethora of other applications, e.g., in more autonomous (without measurement and detailed temporal control) generation of quantum non-Gaussian states for quantum critical sensing \cite{Zana08,Garb20,Chu21,Zhou23,Hott24}, quantum simulations of effects in molecular potentials \cite{OMal16,Shen18,Hu18,Wang20,Dutt24} and, eventually, generation of codes and magic states for quantum computing \cite{Ma21,Eick22,Kudr22,Roda24,Groc23}.

\begin{acknowledgments}
R.F. acknowledges the support of the project no. 22-27431S of the Czech Science Foundation, project CZ.02.01.01/00/22\_008/0004649 (QUEENTEC) of EU and MEYS Czech Republic and the European Union's HORIZON Research and Innovation Actions under Grant Agreement no.\,101080173 (CLUSTEC).
\end{acknowledgments}

\appendix

\section{Effective Hamiltonian}
\label{he}

As outlined in the main text, the effective Hamiltonian of the Rabi model is written in terms of the coordinate and momentum variables in Eq.~\eqref{HamB}, defined through the immediate expectation values of creation and annihilation operators of the oscillator quanta, and by casting the full Hamiltonian in the form proportional to $-\vecb{B}(x,p)\cdot\vecb{\hat{\sigma}}$, where Pauli matrices $\vecb{\hat{\sigma}}\equiv(\hat{\sigma}_x,\hat{\sigma}_y,\hat{\sigma}_z)$ act in the space of the qubit. 
The effective Hamiltonian ${\cal H}_{\mathrm{eff}}(x,p)$ associated with the qubit state $\ket{\!\uparrow_{\vecb{B}}}$ for the quantum Hamiltonian $\hat{H}$ from Eq.~\eqref{Ham} reads as follows: 
\begin{eqnarray}
	\label{eq:Heff}
	\heff(x,p)&\equiv&\frac{{\cal H}_{\mathrm{eff}}(x,p)}{R}=\frac{x^2+p^2}{2}+\sqrt{2}\,\mu\,x\\
	&&-\sqrt{\frac{1}{4}+2(\lambda^2\!+\!\mu^2)\,x^2+2(\lambda\delta)^2\,p^2+\sqrt{2}\,\mu\,x}.\nonumber
\end{eqnarray}
The effective Hamiltonian ${\cal H}'_{\mathrm{eff}}(x,p)$ associated with the opposite qubit state $\ket{\!\downarrow_{\vecb{B}}}$ differs from the one above just by the sign in front of the square-root term ($+$ instead of $-$).
For any finite $R$ these formulas hold only approximately, but their accuracy increases with $R\to\infty$.

\begin{figure}[t]
\includegraphics[width=\linewidth]{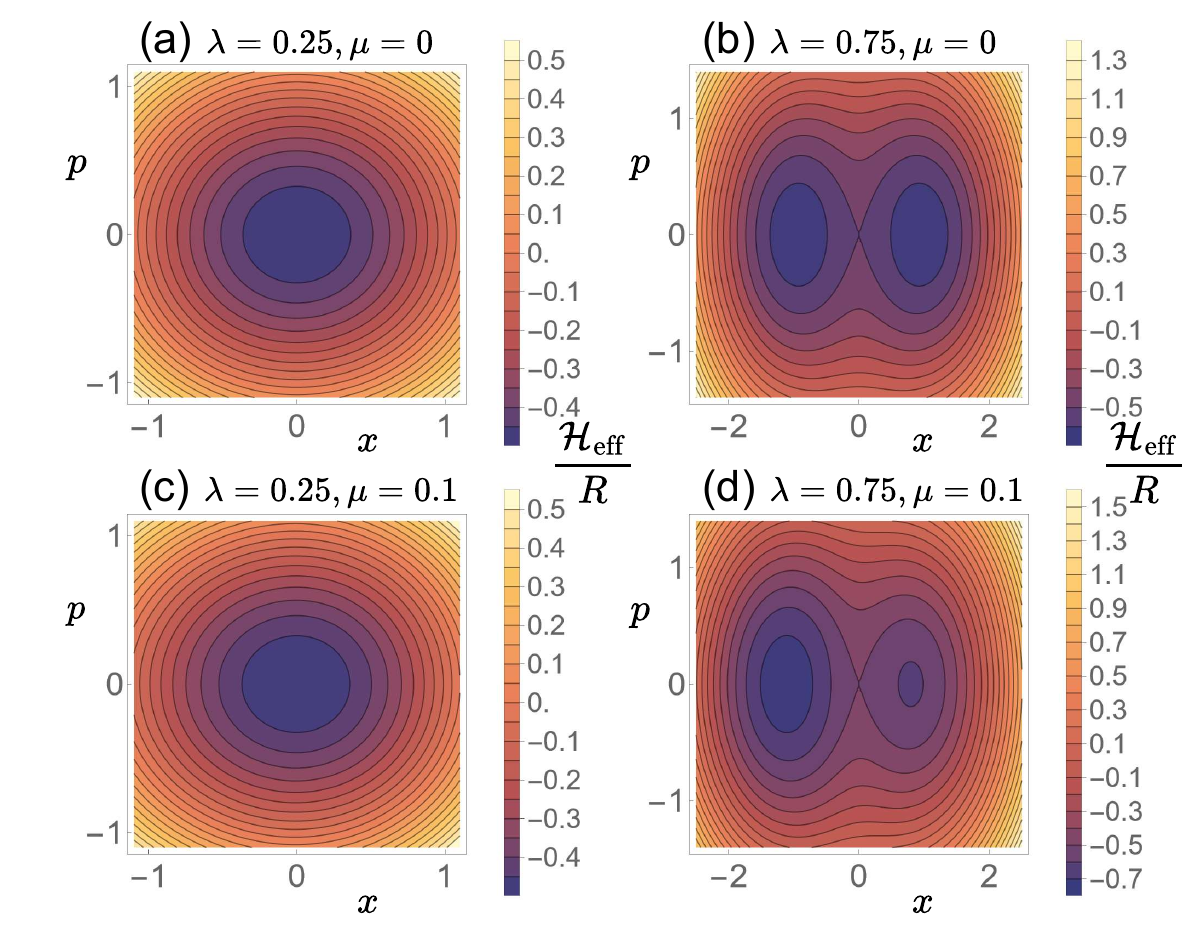}
\caption{The effective Rabi Hamiltonian \eqref{eq:Heff} for various choices of parameters $\lambda$ and $\mu$, with $\delta=0.5$.}
\label{contour}
\end{figure}

Contours of $\heff(x,p)$ from Eq.\,\eqref{eq:Heff} are depicted for ${\delta=0.5}$ (the value set in our analysis) and four choices of interaction strengths $\lambda$ and $\mu$ in Fig.~\ref{contour}.
The upper row shows the single- and double-well forms of $\heff(x,p)$ in the parity conserving, i.e., ${(+x\leftrightarrow-x)}$ symmetric case (${\mu=0}$).
The lower row shows the same in the parity-violating case (${\mu\neq 0}$).
While in panels (a) and (c) the point $(x,p)=(0,0)$ represents the global minimum of $\heff(x,p)$, in panels (b) and (d) it is a saddle point.
For $\lambda>1$ (with the above choice of $\delta$), the $(x,p)=(0,0)$  point would become a~local maximum.
We stress that the value of $\mu$ used in panels (c) and (d) is much larger than the values employed in our analysis of Schr{\"o}dinger-cat states---for the employed values of $\mu$ the reflection asymmetry of contours in Fig.~\ref{contour} would not be visually observable. 

A more comprehensible derivation of the effective Hamiltonian and a more detailed discussion of its forms for various choices of parameter can be found in Ref.\,\cite{Stra21}.

\section{Dynamics near the stationary point and scaling of time}
\label{sc}

Classical dynamics near the stationary point $(x,p)=(0,0)$ is governed by the linearized Hamilton equation
\begin{equation}
	\label{eq:dynam}
	\begin{pmatrix}\dot{x}\\ \dot{p}\end{pmatrix}\approx
	\left.\begin{pmatrix}
	\ \ \partial_{p}\partial_{x}\heff &\ \ \partial_{p}\partial_{p}\heff
	\\
	-\partial_{x}\partial_{x}\heff & -\partial_{x}\partial_{p}\heff,
	\end{pmatrix}\right|_{(x,p)=(0,0)}
	\begin{pmatrix}x\\ p\end{pmatrix} 
\end{equation}
where the dot means the time derivative, and $\partial_{x}$ and $\partial_{p}$  are partial derivatives with respect to variables $x$ and~$p$.
Stability of dynamics near the stationary point depends on the eigenvalues of the $2\times 2$ matrix in Eq.\,\eqref{eq:dynam}:
While real eigenvalues $\Lambda_{\pm}=\pm|\Lambda|$, where $|\Lambda|=|(\partial_{x}\partial_{p}\heff)^2-(\partial_{x}\partial_{x}\heff)(\partial_{p}\partial_{p}\heff)|^{1/2}$,  imply unstable dynamics, imaginary eigenvalues  $\Lambda_{\pm}=\pm i|\Lambda|$ lead to stable dynamics \cite{Pila20,Tabo89}.
In the case of the unstable dynamics we expect that for sufficiently small times $t$ the width $\sigma(t)$ of a state initially centered at the stationary point expands as ${\sigma(t)=\sigma(0)\exp(|\Lambda|t)}$.
In the case of stable dynamics, in contrast, the width $\sigma(t)$ oscillates around the initial value $\sigma(0)$.

Using Eq.\,\eqref{eq:Heff}, we obtain the following formula for eigenvalues of the matrix from Eq.\,\eqref{eq:dynam}: 
\begin{equation}
	\label{eq:eigen}
	\Lambda_{\pm}=\pm\sqrt{(4\lambda^2-1)(1-4\lambda^2\delta^2)}.
\end{equation}
This implies that the dynamics near $(x,p)=(0,0)$ is unstable for $\lambda\in[\frac{1}{2},\frac{1}{2|\delta|}]$ and stable elsewhere.
The unstable case corresponds to the parameter domain where $(x,p)=(0,0)$ is a saddle point of $\heff(x,p)$, whereas the stable case is associated with domains where $(x,p)=(0,0)$ is the global minimum ($\lambda<\frac{1}{2}$) or a local maximum ($\lambda>\frac{1}{2|\delta|}$) of $\heff(x,p)$ \cite{Stra21}.
In the analysis presented in the main text we set $\lambda=0.75$ and $\delta=0.5$, which means that the dynamics around the point $(x,p)=(0,0)$ is unstable with eigenvalues $\Lambda_{\pm}=\pm\sqrt{35}/8=\pm 0.73951$. 

The Wigner distribution $W_{\rm osc}(x,p,0)$ corresponding to the initial vacuum state of the oscillator is the Gaussian of a width $\sigma(0)\propto 1/\sqrt{R}$ centered at $(x,p)=(0,0)$.
Consider two values $R$ and $R'$  satisfying $R<R'$ and two initial states associated with them of widths $\sigma(0)$ and $\sigma'(0)$, respectively.
The time $\Delta t$ that the narrower Gaussian associated with $R'$ needs to expand to the size of the wider Gaussian associated with $R$ is
\begin{equation}
	\label{eq:deltat}
	\Delta t=\frac{1}{|\Lambda|}\ln\frac{\sigma(0)}{\sigma'(0)}=\frac{1}{2|\Lambda|}\ln\frac{R'}{R}.
\end{equation}
After this time, the evolution of both states---in particular their split into two components and travel around both wells of the effective Hamiltonian---is roughly the same.
The delay $\Delta t$ applies again when both components merge after their first revolution and return to the $(x,p)=(0,0)$ saddle point. 
Time relations in the subsequent revolutions basically repeat the scenario of the first one.
Hence the times $\tau$ and $\tau'$ corresponding to the moment of maximal deviation of the average coordinate from zero for both values $R$ and $R'$, i.e., to the first minimum of $\ave{\hat{x}(t')}$ in Fig.~\ref{avex}(a) of the main text, are related by $\tau'=\tau+3\Delta t$, where the time delay $3\Delta t$ counts for two departures from and one arrival to the stationary point.
Therefore, introducing for $R'$ a rescaled time $t'=t/s$ such that the first minimum of $\ave{\hat{x}(t')}$ for~$R'$ coincides with the first minimum of $\ave{\hat{x}(t)}$ for~$R$, we  obtain the following relation for the scaling constant:
\begin{equation}
	\label{eq:s}
	s=\frac{\tau'}{\tau}=1+\frac{3}{2|\Lambda|\tau}\ln\frac{R'}{R}.
\end{equation}

\begin{figure}[t]
	\includegraphics[width=0.95\linewidth]{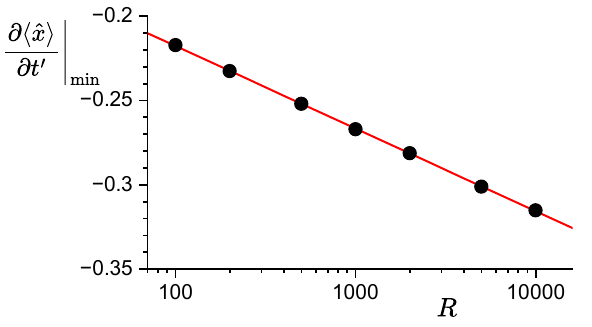}
	\caption{The value $\frac{\partial}{\partial t'}\ave{\hat{x}(t')}$ of the rescaled-time derivative at its first extreme (negative minimum) as a function of the size parameter $R$. 
	Numerical derivatives for 7 values of $R\in[10^2,10^4]$ are shown by black dots. 
	We see that the extremal slope (its absolute value) grows logarithmically, the derivative satisfying $\frac{\partial}{\partial t'}\ave{\hat{x}}\bigr|_{\rm min}\approx -0.049\log R-0.12$ (the red line). Hamiltonian parameters were taken as in the main text: $\lambda=0.75$, $\delta=0.5$. } 
	\label{death}
\end{figure}

Now, let us consider the time derivative $\frac{\partial}{\partial t}$ of the average coordinate $\ave{\hat{x}(t)}$ at the instant of the extremal negative slope of $\ave{\hat{x}(t)}$ after the parity-violating split of the wave function---see Fig.~\ref{avex}, particularly the first minimum of the time derivative in panel (b).
Symbolically, this is the moment when the  Schr{\"o}dinger-cat state dies.
It turns out that the value of the derivative in natural time $t$ is almost independent of $R$, so $\frac{\partial}{\partial t}\ave{\hat{x}(t)}\big|_{\rm min}\approx-v_0$ with $v_0>0$.
Transformation $t\mapsto t'=t/s$  together with formula \eqref{eq:s} leads to
\begin{equation}
	\label{eq:slope}
	\left.\frac{\partial}{\partial t'}\ave{\hat{x}(t')}\right|_{\rm min}\approx-v_0-\frac{3v_0}{2|\Lambda|\tau}\ln\frac{R'}{R},
\end{equation}
which implies a logarithmic decrease of the minimal derivative in the rescaled time with $R$.
	
This is verified in Fig.\,\ref{death}. 
The slope of the red line, which perfectly fits numerical data (the dots), agrees with formula \eqref{eq:slope}, where $\tau$ and $v_0$ are obtained from the numerical simulation of dynamics for $R=10^2$, and $|\Lambda|$ takes the value specified above. 
This means that the absolute value of the slope of $\ave{\hat{x}(t')}$, which quantifies the sharpness of the parity violation effect in the rescaled time, grows logarithmically with the size parameter of the system.
In other words, the death of the Schr{\"o}dinger-cat state in the rescaled time becomes sudden for $R\to\infty$.


\end{document}